\documentclass[aps,twocolumn,prl,superscriptaddress,amsmath,amssymb]{revtex4-1} %% for PRL
\usepackage{dcolumn}
\usepackage{bm}
\usepackage{amsmath}
\usepackage{txfonts}
\usepackage[T1]{fontenc}
\usepackage{xspace}
\usepackage{ulem}
\usepackage{comment}
\ifx\pdfoutput\undefined
\usepackage[dvipdfmx]{graphicx}
\usepackage[dvipdfmx]{hyperref}
\usepackage[dvipdfmx]{color}
\usepackage[dvipdfmx]{xcolor}
\else
\usepackage{graphicx}
\usepackage{hyperref}
\usepackage{color}
\usepackage{xcolor}
\fi
\begin{document}
\let\emph\textit

\title{
  Majorana-mediated spin transport without spin polarization in Kitaev quantum spin liquids
}
\author{Tetsuya Minakawa}
\author{Yuta Murakami}
\author{Akihisa Koga }
\affiliation{
  Department of Physics, Tokyo Institute of Technology,
  Meguro, Tokyo 152- 8551, Japan
}
\author{Joji Nasu}
\affiliation{
  Department of Physics, Yokohama National University, Hodogaya, Yokohama 240-8501, Japan
}

\date{\today}
\begin{abstract}
  We study the spin transport through the quantum spin liquid (QSL)
  by investigating the real-time and real-space dynamics of the Kitaev spin system
  with a zigzag structure in terms of the time-dependent Majorana mean-field theory.
  After the magnetic field pulse is introduced to one of the edges,
  the spin moments are excited in the opposite edge region
  although no spin moments are induced in the Kitaev QSL region.
  This unusual spin transport originates from the fact that
  the $S=1/2$ spins are fractionalized into the itinerant and localized Majorana fermions
  in the Kitaev system.
  Although both Majorana fermions are excited by the magnetic pulse,
  only the itinerant Majorana fermions flow through the bulk regime
  without the spin excitation, resulting in the spin transport in the Kitaev system.
  We also demonstrate that this phenomenon can be observed even in the system
  with the Heisenberg interactions using the exact diagonalization.
\end{abstract}
\maketitle

Spin transport without an electric current has attracted not only
practical interest in spintronics but also considerable attention
in modern condensed matter physics.
In insulating magnets, the carriers of the spin current are conventionally considered
to be magnons, which are elementary excitations
in a magnetically ordered state~\cite{uchida2010,uchida2010spin,Xiao2010,Rezende2014}.
By contrast, the possibility of the spin transport
in the quantum spin liquid (QSL)
has been discussed recently.
One of the typical examples is an antiferromagnetic Heisenberg spin-1/2 chain,
where elementary excitations are described by the spinon with an $S=1/2$ spin.
The spin Seebeck experiments for the cuprate $\rm Sr_2CuO_3$
have clarified that the spin current arises even in the QSL~\cite{hirobe2017one}.
Therefore, the spinons, instead of the magnons, can be responsible
for the spin transport in the nonmagnetic system.

Another interesting playground % for the spin transport in
of the QSL is the Kitaev model~\cite{KITAEV2006},
which has been studied intensively in this decade~\cite{PhysRevLett.100.177204,Jackeli_2009,Chaloupka_2010,PhysRevLett.105.117201,PhysRevB.84.165414,Chaloupka_2013,Yamaji_2014,Nasu1,Nussinov,Suzuki_2015,PhysRevB.92.075114,PhysRevX.5.041035,Yamaji_2016,PhysRevLett.117.277202,PhysRevLett.117.037209,yadav2016kitaev,Trebst,Winter_2017,Gohlke_2017,Nasu2,Hermanns,Tomishige_2018,Nakauchi_2018,Nasu_2018,KnolleMoessner,Takagi_2019,PhysRevLett.123.087203,MotomeNasu}.
The Kitaev model consists of bond-dependent Ising interactions
between spin-1/2 moments on a honeycomb lattice,
and its ground state is exactly shown to be a QSL.
One of the interesting features is the spin fractionalization.
Namely,
the spins are fractionalized into itinerant and
localized Majorana fermions.
Since both quasiparticles are charge neutral,
the thermal transport is one of the most promising phenomena
to grasp the presence of the Majorana fermions.
Particularly, a half quantized plateau in the thermal quantum Hall effects
has been successfully observed in the Kitaev candidate material
$\alpha$-RuCl$_3$~\cite{Plumb2014,Kubota2015,Sears2015,Majumder2015},
which is a direct evidence of a topologically protected chiral
Majorana edge mode~\cite{kasahara2018majorana}.
On the other hand, less is known about
the Majorana-mediated spin transport in the Kitaev QSL
although it has recently been discussed in the related systems~\cite{Hong,Carvalho,Aftergood}.

In the Kitaev model, spin correlations are extremely short-ranged
due to the existence of the local $Z_2$ symmetry,
in contrast to the Heisenberg chain with power-low spin correlations.
However, it does not necessarily mean the absence of
the spin transport in the Kitaev model.
When small local perturbations are present in the system,
{\it eg.} the magnetic field, edges, defects, etc.
the $Z_2$ symmetry is lost in certain regions~\cite{PhysRevB.99.184418}.
Therefore, intriguing phenomena are expected to be induced in these regions.
For example, the spin excitation could flow through the Kitaev QSL region without spin polarization.
Thus, it is highly desired to examine the spin transport in the nonequilibrium dynamics,
which should be important to observe the itinerant nature of the Majorana fermions
in the bulk.

In this Letter,
to address the spin transport through the Kitaev QSL,
we investigate the real-time dynamics triggered by an impulse magnetic field
on one of the edges.
Using the time-dependent mean-field (MF) theory,
we examine the time evolution of the magnetization and dynamics of the fractionalized Majorana quasiparticles.
We demonstrate that a spin-polarized wavepacket created at the edge propagates to the other edge even when the two edges are separated by the QSL region without spin polarization.
We also address how robust this anomalous phenomenon is against the
Heisenberg interactions by means of the exact diagonalization (ED).
Finally, we propose the ways to extract the results intrinsic to the Kitaev QSL with the fractionalized quasiparticles in experiments.

\begin{figure}[t]
\includegraphics[width=\columnwidth,clip]{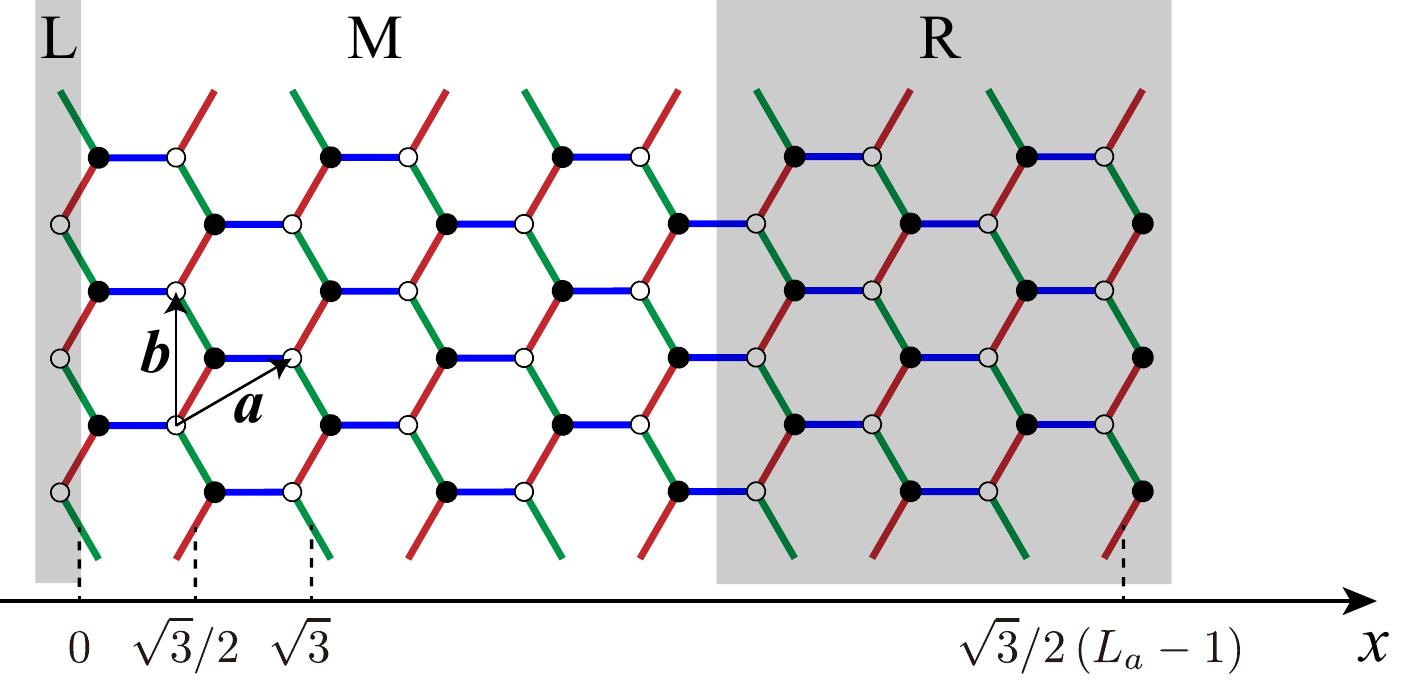}
\caption{ Kitaev model on a honeycomb lattice with zigzag edges.
  Red, blue, and green lines represent $x, y$, and $z$ bonds, respectively.
  Solid (open) circles represent spin-1/2 in the $A$ ($B$) sublattice.
In this figure, four $z$ (green) bonds exist along the $\bm{a}$ direction in the R region, namely, $L_R=4$.}
\label{fig:model}
\end{figure}

We consider the Kitaev model in the $L_a\times L_b$ cluster of the honeycomb lattice
with zigzag edges, which is schematically shown in Fig.~\ref{fig:model}.
The norms of the primitive translational vectors $\bm{a}$ and $\bm{b}$ are assumed to be unity.
The periodic boundary condition is imposed along the $\bm{b}$-direction.
The system we consider here is composed of three regions.
In the middle (M) region, no magnetic field is applied and
the Kitaev QSL is realized without spin polarization.
In the right (R) region, the static magnetic field $h_R$ is applied.
We introduce $L_R$, which is defined as the number of $z$ bonds included in this region with respect to the $\bm{a}$ direction (see Fig.~\ref{fig:model}).
Moreover, we term the L region composed of the left-edge sites.
In this region, we introduce the time-dependent magnetic field $h_L(t)$.
The corresponding Hamiltonian is
\begin{eqnarray}
  {\mathcal H}(t)= -J_K\sum_{\gamma=x,y,z}\sum_{\langle i,j \rangle_\gamma}S_i^\gamma S_j^\gamma - h_R\sum_{i\in {\rm R}} S_i^z  - h_L(t)\sum_{i\in {\rm L}} S_i^z,
\label{eq:Model}
\end{eqnarray}
where $S_i^\gamma$ is the $\gamma(=x,y,z)$ component of an $S=1/2$ spin operator
at the $i$th site.
The ferromagnetic exchange $J_K(>0)$ is defined on three different types of
the nearest-neighbor bonds, $x$ (red), $y$ (blue), and $z$ (green) bonds (see Fig.~\ref{fig:model}).
%%%%%%%%%%%%%%%%%%%%%%%%%%%%%%%%%%%%%
%%%%%%%%%%%%%%%%%%%%%%%%%%%%%%%%%%%%%
It is known that, in the uniform lattice,
the magnetic field induces the phase transition
to the spin-polarized state around $h_c/J_K\sim 0.042$~\cite{Nasu_2018} within the MF theory.
Therefore, we restrict ourselves to the case with $h_R<h_c$
to discuss the spin transport inherent in the Kitaev QSL.

We study the time evolution of the system
upon stimuli of the magnetic pulse
in the L region (see Fig.~\ref{fig:model})~\cite{Misawa2019,Misawa2019pre}.
To this end, we introduce the time-dependent Majorana MF
theory.
The details of the formulations are shown in Ref.~\cite{Nasu_quench2019}.
The Hamiltonian Eq.~(\ref{eq:Model}) is obtained as a fermion model
by applying the Jordan-Wigner transformation to the spin operators~\cite{PhysRevB.76.193101,PhysRevLett.98.087204,1751-8121-41-7-075001}.
Furthermore, by introducing two kinds of Majorana fermions,
$\gamma$ and $\bar{\gamma}$, from a complex fermion at each site,
Eq.~(\ref{eq:Model}) is rewritten as
\begin{eqnarray}
{\cal H}(t)&=&-\frac{J_K}{4}\sum_{\bm{r}}\left(i\gamma^A_{\bm{r}-\bm{a}+\bm{b}}\gamma^B_{\bm{r}}+i\gamma^A_{\bm{r}+\bm{b}}\gamma^B_{\bm{r}} \right)
-\frac{J_K}{4}\sum_{\bm{r}} i\gamma^A_{\bm{r}}\gamma^B_{\bm{r}} i\bar{\gamma}^A_{\bm{r}}\bar{\gamma}^B_{\bm{r}}\nonumber\\
&-&\frac{h_R}{2}
\sum_{\bm{r}\in \textrm{R}}\left(i\gamma^A_{\bm{r}}\bar{\gamma}^A_{\bm{r}}-i\gamma^B_{\bm{r}}\bar{\gamma}^B_{\bm{r}} \right)
+\frac{h_L(t)}{2}\sum_{\bm{r}\in \textrm{L}} i \gamma^B_{\bm{r}}\bar{\gamma}^B_{\bm{r}},
\end{eqnarray}
where ${\bm r}$ indicates the position of the $z$ bond (the center of the $z$ bond; see Fig.~\ref{fig:model}).
$\gamma_{\bm{r}}^A$ and $\bar{\gamma}_{\bm{r}}^A$ ($\gamma_{\bm{r}}^B$ and $\bar{\gamma}_{\bm{r}}^B$)
are the Majorana fermion operators
 connected with the $z$ bond
 in the sublattice $A$ ($B$), as shown in Fig.~\ref{fig:model}.
When $h_R=h_L(t)=0$,
$[{\mathcal H}, \eta_{\bm r}]=0$ at each $\bm{r}$,
and $\eta_{\bm r}(=i\bar{\gamma}^A_{\bm{r}}\bar{\gamma}^B_{\bm{r}})$ is the $Z_2$ local conserved quantity.
In the case, the model is solvable
as the Hamiltonian is bilinear in terms of $\gamma$, and its low-energy dispersion is given as
$\varepsilon_{\bm{k}}\simeq v |\bm{k}-\bm{k}_{\textrm{K}}|$
around the K point with the velocity $v=\sqrt{3}J_K/4$.
This indicates that $\gamma$ and $\bar{\gamma}$ are regarded as
the itinerant and localized Majorana fermions, respectively.

Since the magnetic field hybridizes two kinds of the Majorana fermions,
the Hamiltonian is no longer exactly solvable.
Here, we apply the Hartree-Fock type decoupling to the interaction on the $z$ bond as
$i\gamma^A_{\bm{r}}\gamma^B_{\bm{r}} i\bar{\gamma}^A_{\bm{r}}\bar{\gamma}^B_{\bm{r}}
 \sim i\gamma^A_{\bm{r}}\gamma^B_{\bm{r}} \Theta_1(x,t) +\Theta_2(x,t)  i\bar{\gamma}^A_{\bm{r}}\bar{\gamma}^B_{\bm{r}} - \Theta_1(x,t)\Theta_2(x,t) -  i\gamma^A_{\bm{r}}\bar{\gamma}^A_{\bm{r}} \Theta_3(x,t) -\Theta_4(x,t)  i \gamma^B_{\bm{r}}\bar{\gamma}^B_{\bm{r}} + \Theta_3(x,t)\Theta_4(x,t)  - i\gamma^A_{\bm{r}}\bar{\gamma}^B_{\bm{r}} \Theta_5(x,t) -\Theta_6(x,t)  i\bar{\gamma}^A_{\bm{r}}\gamma^B_{\bm{r}} + \Theta_5(x,t)\Theta_6(x,t)$,
where we have introduced the six kinds of
$x$- and $t$-dependent MFs as
$\Theta_1(x,t) = \langle i\bar{\gamma}^A_{\bm{r}}\bar{\gamma}^B_{\bm{r}} \rangle \equiv \langle \eta \rangle(x,t)$,
$\Theta_2(x,t) = \langle  i\gamma^A_{\bm{r}}\gamma^B_{\bm{r}} \rangle  \equiv  \langle \xi \rangle(x,t)$,
$\Theta_3(x,t) = \langle  i\gamma^B_{\bm{r}}\bar{\gamma}^B_{\bm{r}} \rangle  = -2\langle S_{B}^z\rangle(x,t)=-2\langle S^z\rangle(x-x_d/2,t)$,
$\Theta_4(x,t) = \langle  i\gamma^A_{\bm{r}}\bar{\gamma}^A_{\bm{r}} \rangle  = 2\langle S_{A}^z\rangle(x,t)=2\langle S^z\rangle(x+x_d/2,t)$,
$\Theta_5(x,t) = \langle  i\bar{\gamma}^A_{\bm{r}}\gamma^B_{\bm{r}} \rangle$,
$\Theta_6(x,t) = \langle  i\gamma^A_{\bm{r}}\bar{\gamma}^B_{\bm{r}}  \rangle$,
where $x$ is the horizontal coordinate of $\bm{r}$ and $x_d=1/(2\sqrt{3})$.
This MF theory is exact when $h_L=h_R=0$.
Therefore, we believe that our MF results are reliable
as far as the small fields are applied.
In the MF theory, the many-body wave function is expressed as a direct product of one-body states,
whose time-evolution is described by the MF Hamiltonian.
Determining the MFs at each time from the many-body wave function, we compute the 
time-evolution of the one-body states with the extended Euler method~\cite{terai1993solitons,Hirano2000,Tanaka2010,Ohara_2017,Tanaka2018,Seo2018}.
The time-dependent magnetic field is explicitly given as,
%%%%%%%%%%%%%%%%%%%%%%%%%%%%%%%%%
  $h_L(t) = \frac{A}{\sqrt{2\pi}\sigma}\exp
  \left[-\frac{t^2}{2\sigma^2}\right]$,
where $A$ and $\sigma$ are constants for the Gaussian pulse.
In the following,
we fix the system size as $L_a=50$ and $L_b=300$, the static field as $h_R = 0.01J_K$, and pulse parameters as {\bf $A=1$} and $\sigma=2J_K^{-1}$.

\begin{figure}[t]
\includegraphics[width=\columnwidth,clip]{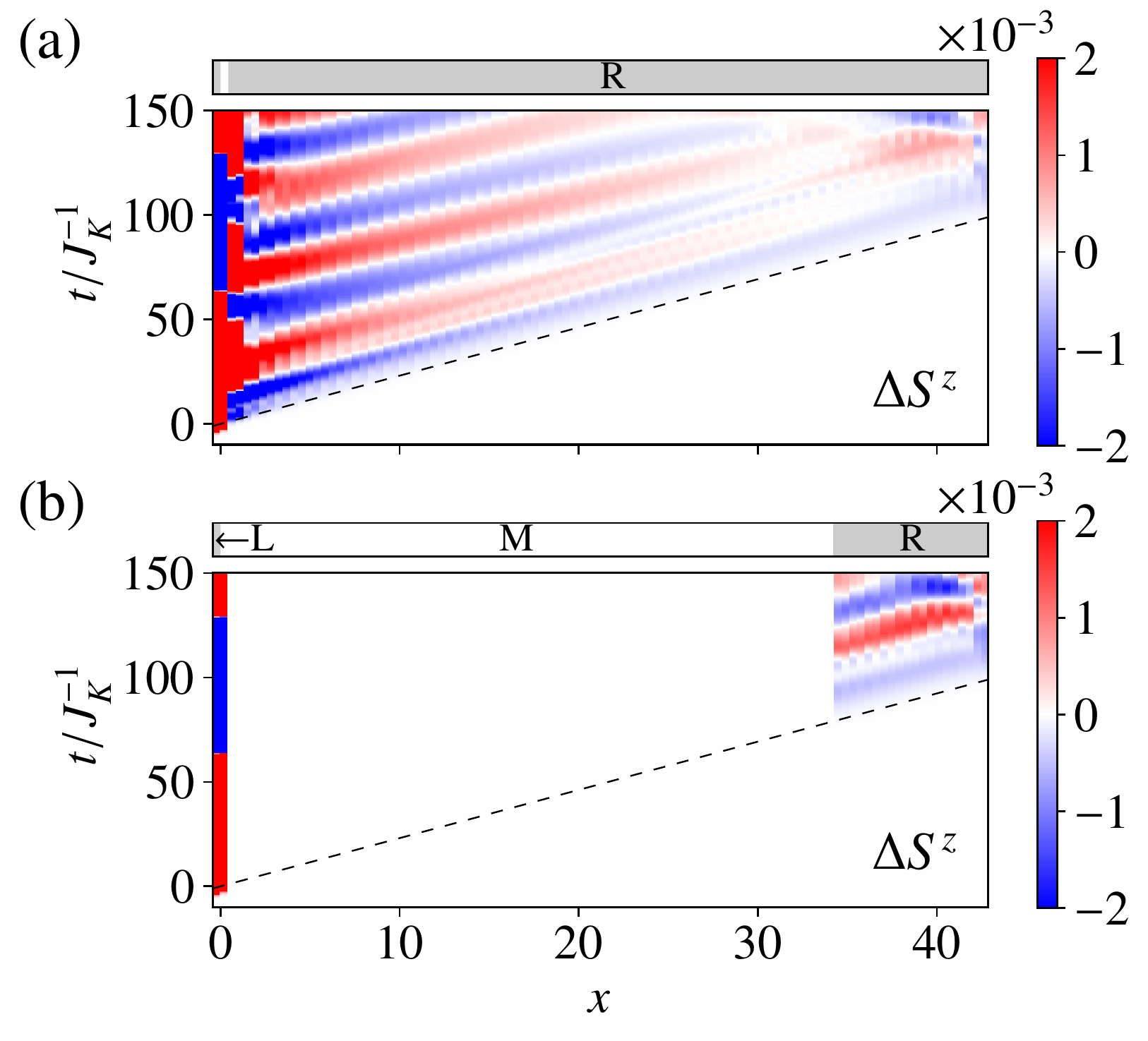}
\caption{   Real-time dynamics of the Kitaev spin system with $L_a=50$ and $L_b=300$
  after the magnetic-field pulse at $t=0$.
  The contour plot of $\Delta S^z(x,t)$ on the plane of the time $t$ and the space $x$ in the system (a) without and (b) with the M (white) region where the QSL is realized with no spin polarization between the L (left gray) and R (right gray) regions (see the top of the panels).
The dashed lines represent $x=vt$ with the Majorana velocity $v$ (see text).}
\label{fig:2}
\end{figure}

Before discussing the spin transport through the M region,
we examine the system without this region, namely,
the static magnetic field $h_R$ is applied
to the sites in the R region with $L_R=L_a-1$.
In this case, there are no local conserved quantities,
leading to nonzero local spin moments.
Figure~\ref{fig:2}(a) shows the contour plot of the changes
in the spin moments $\Delta S^z(x,t)$, where $\Delta O(x,t)=\langle O(x,t)\rangle-\langle O(x,-\infty)\rangle$.
As expected, we find that the wavepacket created by the magnetic-field pulse flows to the right edge.
Note that its velocity almost coincides with the Majorana velocity of
the genuine Kitaev model, $v$, which is shown as the dashed line in Fig.~\ref{fig:2}(a).
This indicates that the propagation is attributed to the gapless Majorana excitation in the bulk within a small static field.

%%%%%%%%%%%%%%%%%%%%%%%%%%%%%%%%%%%%%%%%%%%%%%%%%%%%%
%%%%%%%%%%%%%%%%%%%%%%%%%%%%%%%%%%%%%%%%%%%%%%%%%%%%%%

Now, we consider the real-time dynamics of the nonmagnetic Kitaev spin system triggered
by the magnetic-field pulse at the left edge
to discuss how the wavepacket flows through the M region (the Kitaev QSL without spin polarization).
In this region, the local conserved quantity is present in each hexagon,
and spin correlations are extremely short-ranged~\cite{Baskaran2007}.
Figure~\ref{fig:2}(b) shows the time evolution of
$\Delta S^z(x,t)$ in the system
with $L_R=10$.
 We find that the magnetic moment is always zero
 in the M region
and no proximity effect is found around the interface between L and M regions.
Nevertheless, in the R region, $\Delta S^z(x,t)$ is induced and the wavepacket flows with the Majorana velocity $v$.
This result indicates that the spin excitations propagate in the nonmagnetic region via the itinerant Majorana fermions, which cannot be explained by classical pictures such as the spin wave theory.

%%%%%%%%%%%%%%%%%%%%%%%%%%%%%%%%%%%%%%%%%%%%%%%%%%%%%%
  \begin{figure}[t]
\includegraphics[width=\columnwidth,clip]{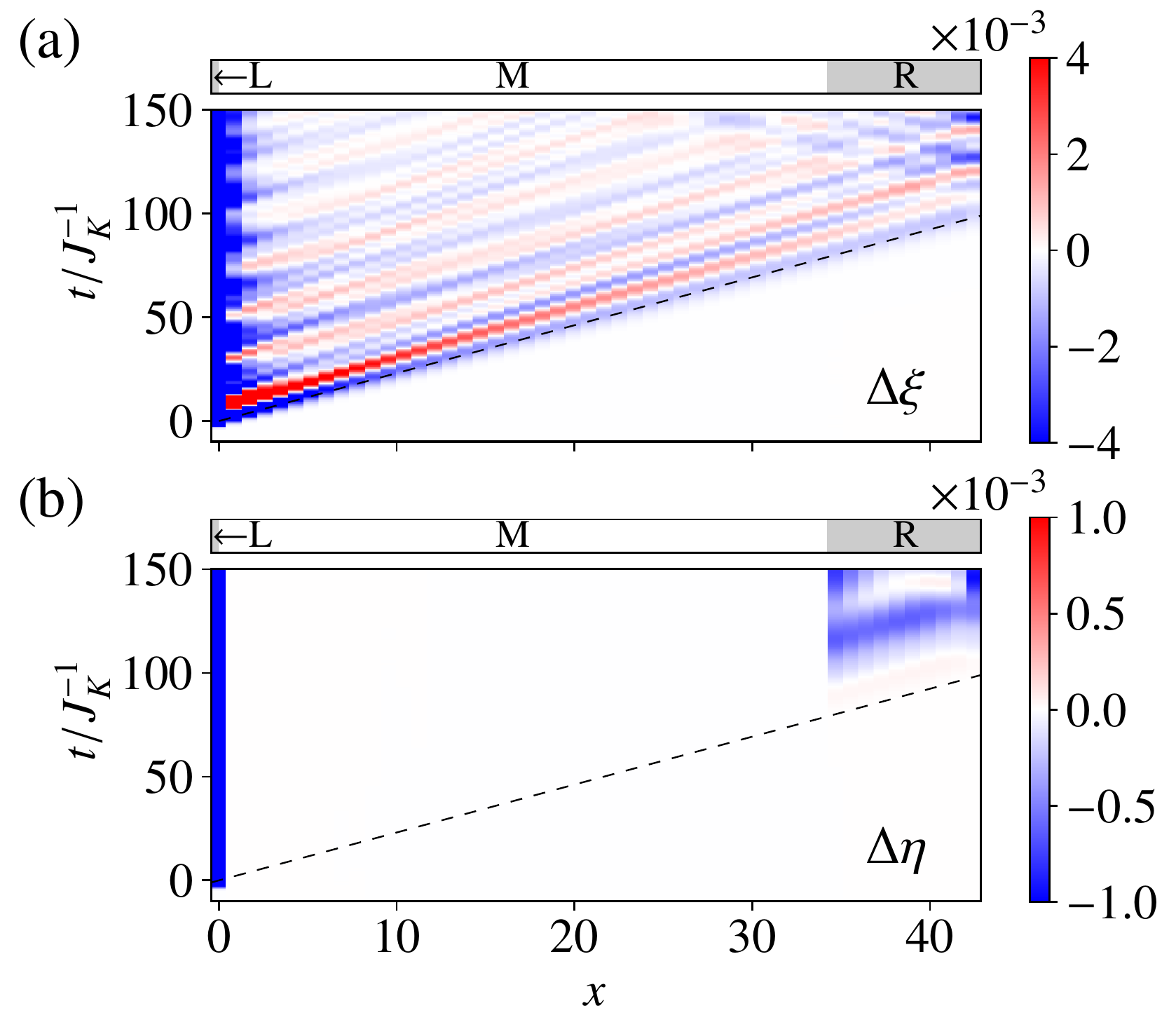}
\caption{     Real-time evolution of (a) $\Delta\xi$, and (b) $\Delta\eta$ of the Kitaev spin system with $L_a=50$ and $L_b=300$
  after the magnetic-field pulse at $t=0$.  The setup and parameters are the same as in Fig.~2(b).}
\label{fig:3}
\end{figure}
%%%%%%%%%%%%%%%%%%%%%%%%%%%%%%%%%%%%%%%%%%%%%%%%%%%%%
To discuss the propagation of the spin excitation through the QSL region
in more detail,
we examine the time evolutions of $\Delta \xi$ and $\Delta \eta$,
which correspond to the dynamics of itinerant and localized Majorana fermions,
as shown in Figs.~\ref{fig:3}(a) and \ref{fig:3}(b).
We find in Fig.~\ref{fig:3}(a) that
the excitation created in the left edge at $t=0$ propagates in the whole region,
which results from the motion of the itinerant Majorana fermions. By contrast, Fig.~\ref{fig:3}(b) shows that $\Delta \eta$ vanishes
in the M region
owing to the existence of the local $Z_2$ symmetry,
while it appears in the R region.
This is the similar behavior as in $\Delta S^z$.
These suggest that,
after the excitation at the left edge, only the itinerant Majorana fermions propagate
in the bulk, where no oscillation appears in the magnetization,
and finally reach the R region.
The weak magnetic field in the R region yields
the hybridization between the itinerant Majorana fermions and the localized fermions,
resulting in time-dependent nonzero spin moments there.
Thus, in the Kitaev QSL,
the spin transport is mediated by the Majorana fermions
although the spin moments never appear.
Moreover, we have confirmed that the magnitude of the spin moment induced in the R region
exhibits a power-low decay as a function of the length of the M region.
This is ascribed to the gapless dispersion of the Majorana fermions,
in contrast to the existence of the gap in the spin excitation in the Kitaev model.

The pulse-amplitude dependence in this phenomenon is also remarkable.
In the R region, $\Delta S^z$ turns out to be proportional to $ A^2$.
This can be explained by considering the local symmetry
at the left edge~\cite{proof}.
This non-linear feature is intrinsic in the Kitaev model,
in contrast to the conventional systems with $\Delta S^z \propto A$.
To study how visible anomalous behavior is in the system with the Heisenberg interaction,
we apply the ED method to the Hamiltonian
${\cal H}(t)+J_H\sum_{\langle i,j \rangle}\bm{S}_i\cdot \bm{S}_j$ with
the antiferromagnetic Heisenberg coupling $J_H(>0)$.
It is known that, when $h_R=h_L(t)=0$, the Kitaev QSL is stable
against small $J_H$~\cite{Chaloupka_2010,Chaloupka_2013,Yamaji_2016,Gohlke_2017}.
In our calculations, the initial ground state is obtained with the Lanczos method and
the time evolution is simply evaluated by the Runge-Kutta method.

The obtained results for the 24-site cluster with $L_a=4$, $L_b=3$, and $L_R=1$ are shown in Fig.~\ref{fig:3}.
In the calculations,
we have confirmed that the induced moment is always parallel to the $z$ direction.
  \begin{figure}[t]
\includegraphics[width=\columnwidth,clip]{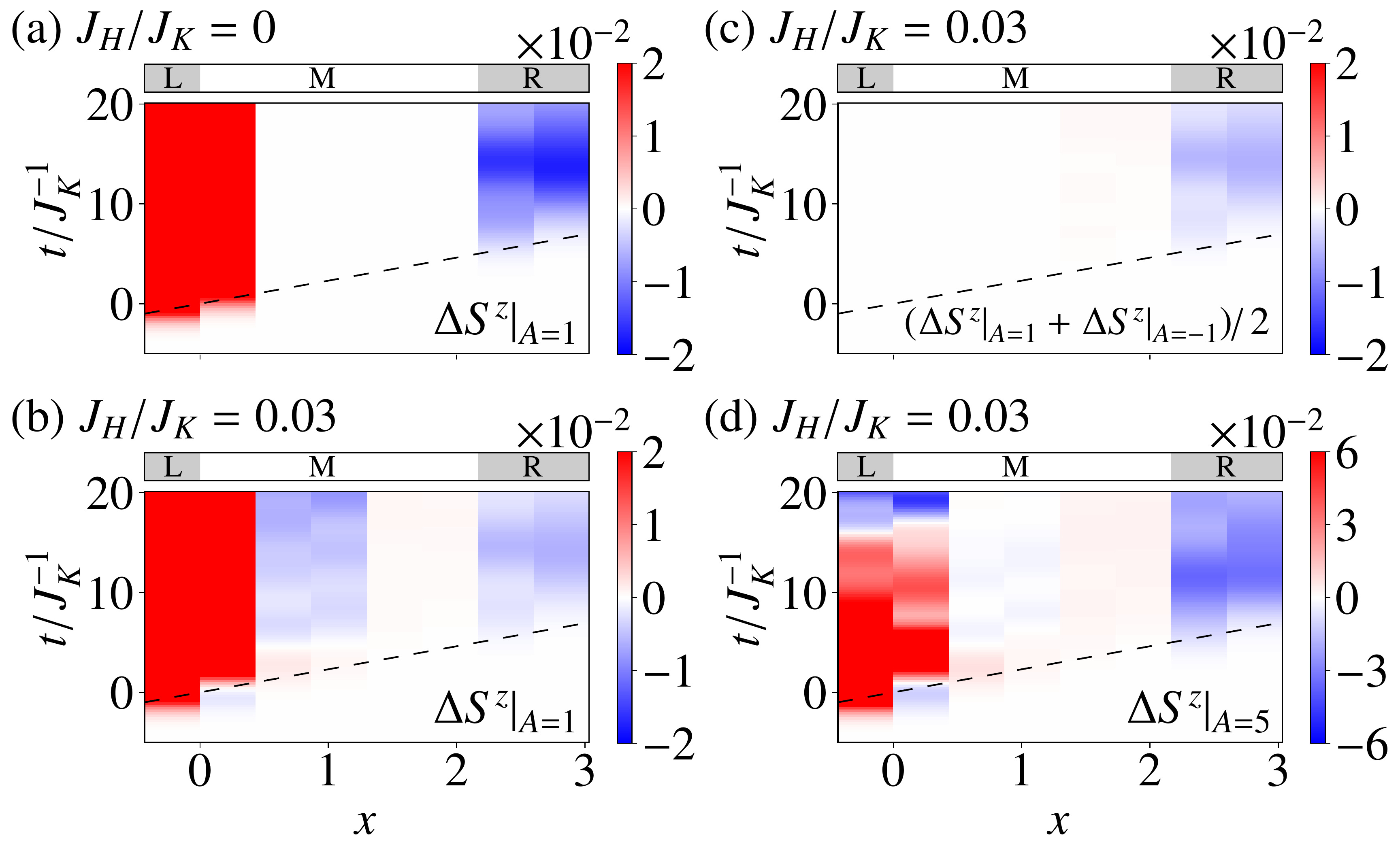}
\caption{    Contour plots of time-dependent spin moment $\Delta S^z(x,t)$
  in the 24-site cluster with $h_R = 0.01 J_K$
  when (a) $(J_H/J_K,A)= (0,1)$, (b) $(0.03,1)$ and (d) $(0.03,5)$.
(c) Contour plot for the average of $\Delta S^z(x,t)$ with $(J_H/J_K,A)= (0.03,1)$ and $(0.03,-1)$.
The dashed lines represent $x=vt$ with the Majorana velocity $v$ (see text).}
\label{fig:4}
\end{figure}
First, we show the results for the genuine Kitaev model with $J_H=0$
in Fig.~\ref{fig:4}(a).
One can find the propagation of the magnetic excitation from one edge to the other
through the QSL region without spin polarization, which is consistent with the Majorana MF result discussed above.
In the presence of the Heisenberg term ($J_H/J_K=0.03$),
$\Delta S^z(x,t)$ takes nonzero values in the M region, as shown in Fig.~\ref{fig:4}(b),
suggesting that the Heisenberg interaction affects
the flow of the spin excitation.
In particular, the spin modulation in the M region is more prominent compared to that in the R region.
This difference from the genuine Kitaev model originates from
the fact that the Heisenberg interaction yields
the interaction between itinerant and localized Majorana fermions.
Therefore, the spin moments appear in the M region
near the interface to the L region as a proximity effect, as shown in Fig.~\ref{fig:4}(b).

We note that
$\Delta S^z$ in the R region is similar to the case
without the Heisenberg interactions.
This implies that
the spin transport inherent in the Kitaev model still survives. % in the R region.
It is naively expected that in the M and R region, the Heisenberg and Kitaev interactions mainly give $A$ and $A^2$ contributions in the spin oscillation, as discussed above. Therefore,
the unique feature for the Kitaev system is extracted
by examining the average of the magnetic responses
after the magnetic pulses with $A$ and $-A$.
In Fig.~\ref{fig:4}(c), this quantity is hardly seen in the M region but clearly observed in the R region, which is a consequence of the Kitaev QSL with itinerant Majorana fermions.

When the pulse amplitude $A$ is relatively large,
the Kitaev interaction plays a dominant role for the spin propagation and
the spin transport without spin polarization becomes practically prominent.
Figure~\ref{fig:4}(d) presents the results with the large $A$.
The spin moments induced in the M region are relatively small, but the spin excitation propagates to the right edge, at which
the spin moments induced are much larger than those in the M region.
This phenomenon is essentially the same as that
in the genuine Kitaev case shown in Fig.~\ref{fig:4}(a).
The above two results suggest that
the spin transport mediated by the fractionalized itinerant quasiparticles
without spin excitations can be observed
even in the presence of additional interactions.

Finally, we discuss the relevance of the present results to real materials.
The setup of our study could be implemented by considering a Kitaev candidate material sandwiched by ferromagnetic insulators.
The candidate materials have been proposed as
$A$IrO$_3$ (A=Na, Li)~\cite{PhysRevB.82.064412,PhysRevLett.108.127203,PhysRevLett.109.266406,PhysRevLett.108.127204,PhysRevLett.114.077202,Kitagawa2018nature} and
$\alpha$-RuCl$_3$~\cite{Plumb2014,Kubota2015,Sears2015,Majumder2015}.
The stimuli of the magnetic field pulse can be injected from a ferromagnetic insulator by the spin pumping~\cite{kajiwara2010transmission,Sandweg2011,Heinrich2011,Hahn2013} or circular polarized light irradiation~\cite{kimel2005ultrafast,Stanciu2007}.
Our results suggest that the spin-excited flow propagates to the other edge
even if the magnetic polarization is absent in the Kitaev magnet, and therefore, we expect that the time-dependent magnetic moment is observed in the ferromagnetic insulator connected to the other side of the Kitaev magnet with a small overlapping.
This time evolution can be experimentally measured by the Kerr or Faraday rotations~\cite{Hiebert1997,kimel2005ultrafast}, which will provide convincing evidence of the fractionalized itinerant quasiparticles in the bulk of the Kitaev magnet.

Note that in the real system, a magnetic order hinders the appearance of the Kitaev QSL~\cite{PhysRevB.83.220403,PhysRevB.85.180403,PhysRevB.92.235119,PhysRevB.93.134423,freund2016single,PhysRevB.93.195158}.
This effect can be avoided by the finite temperature measurement above the N\'eel temperature, where the itinerant quasiparticles are active, and/or the recent progress of the thin film~\cite{Yamaji_2014,Winter2016,weber2016magnetic,ziatdinov2016atomic,gronke2018chemical,zhou2019possible,Zhou2019,mashhadi2019spin,Gerber2019pre,Biswas2019}, which suppresses the magnetic ordering due to the suppression of the interlayer coupling.
Moreover, by changing the intensity of the injection of the spin excitation, one could estimate the magnitude of the additional interactions such the Heisenberg one.
The effect of the off-diagonal interactions, so called $\Gamma$ term, is not addressed in the present study but we expect that this gives a similar effect to the Heisenberg one~\cite{PhysRevLett.112.077204,PhysRevB.99.064425,gordon2019theory}.

In summary, we have demonstrated that, after the magnetic excitation at one of the edges in the Kitaev spin system,
the spin moments never appear in the bulk, but
are fluctuated in the opposite edge.
We have revealed that this unusual spin transport is governed by the fractionalized itinerant Majorana fermions.
The spin transport without spin polarization should be visible
even in the system with the Heisenberg coupling
by using the pulse field dependence in $\Delta S^z$.

We also note that it might be possible to control the motion of
the localized Majorana fermions (vison)
in the bulk, by switching on/off the magnetic field. % (not shown)
This should be important for realizing the vison transport in the experiments.
It is also interesting to study the spin transport
in the generalized Kitaev models
~\cite{PhysRevB.78.115116,S1Koga,Minakawa}, where the existence of spin fractionalization has also been suggested~\cite{S1Koga,Oitmaa,MixedKoga}.
The real-time spin dynamics should be one of the possible candidates
to clarify the presence of the quasiparticles.

\begin{acknowledgments}
The authors thank T. Mizoguchi and M. Udagawa for fruitful discussions.
Parts of the numerical calculations were performed in the supercomputing
systems in ISSP, the University of Tokyo.
  This work was supported by Grant-in-Aid for Scientific Research from
  JSPS, KAKENHI Grant Nos. JP19K23425 (Y.M.),
  JP19H05821, JP18K04678, JP17K05536 (A.K.),
JP16H02206, JP18H04223, JP19K03742 (J.N.), by JST CREST Grant No. JPMJCR1901 (Y.M.) and by JST PREST (JPMJPR19L5) (J.N.).
\end{acknowledgments}

\bibliography{./refs}

\begin{thebibliography}{93}
\bibitem [{pro()}]{proof}%
  \BibitemOpen
  \href@noop {} {}\bibinfo {note} {This can be explained as follows. Since the
  system has the local symmetry before and after the introduction of the pulsed
  field, each eigenstate is specified by a set of the eigenvalues of $\eta$ in
  the L region. Let us start with the initial state given as $|\Psi;\eta_1,
  \eta_2, \cdots\rangle$. After the pulse at the left-edge, the final state
  should be given within the first-order perturbation as
  $|\Psi(t)\rangle=|\Psi';\eta_1, \eta_2, \cdots\rangle + A\sum_i c_i |\Phi_i;
  \eta_1,\eta_2,\cdots, -\eta_i,\cdots, \rangle$, where $c_i$ is some constant
  and $|\Psi'\rangle$ and $|\Phi_i\rangle$ belong to the distinct subspaces
  specified by different sets of eigenvalues of $\eta$ at the left edge. On the
  other hand, when some operator $O$ commutes with $\eta$ at the left edge,
  $O|\Psi'\rangle$ belongs to the same subspace as $|\Psi'\rangle$. (For
  example, $O$ is $S_z,\xi$,or $\eta$ in the M and R regimes.) Therefore, the
  first order component of $A$ in $\langle \Psi(t)| O|\Psi(t)\rangle$
  vanishes}\BibitemShut {NoStop}%
\end{thebibliography}

\end{document}